\documentclass[10pt]{article}

\usepackage{latexsym}

\usepackage{amsmath}

\usepackage{graphicx}

\begin{document}

%%%%%%%%%%%%%%%%%%%%%%%%%%%%%%%%%%%%%%%%%%%%%%%%%%%%%%%%%%%%%%%%%%%

\title{ Self-consistency functional equation in strongly correlated electron systems: a different approach}

\author{
     Ekaterina Grancharova \thanks{E-mail: granch@phys.uni-sofia.bg}\\
{\footnotesize  Department of Semiconductor Physics,
                Faculty of Physics, Sofia University,}\\
{\footnotesize  5 James Bourchier Boulevard, Sofia~1164, Bulgaria }\\
}

\date{25 March 2006}

\maketitle

\begin{abstract}
We propose a new approach to the self-consistency equation, which arises in the problem of the motion 
of a hole in a quantum antiferromagnet, appropriate to the case of small exchange energy $J$. The functional 
equation for the Green function is transformed into a differential equation; its solutions are analyzed and compared 
to the existing numerical calculations. This method allows one to study the limit of $J\to 0$. Application to other strongly 
correlated electron systems is discussed.
\end{abstract}

%%%%%%%%%%%%%%%%%%%%%%%%%%%%%%%%%%%%%%%%%%%%%%%%%%%%%%%%%%%%%%%%%%%

%\draft
\sloppy

\section{Introduction}

The interplay between doping and antiferromagnetism is believed to play an important role in high $T_c$ superconductors. Hence the motion of holes 
in an antiferromagnetic background has received much attention in the last two decades.
After the pioneering work of Brinkman and Rice \cite{BR}, where the Hubbard model with $U=\infty$ is considered, many publications on this subject are 
known \cite{Mart}, \cite{SR}, \cite{Szcz}, \cite{Kane}.
 Often this problem is faced by applying the model t-J Hamiltonian, acting on the space with no doubly occupied sites, which is obtained from the 
Hubbard  large U model. One may use, for instance, the Holstein-Primakoff transformation (see \cite{SR}) to rewrite  the Hamiltonian it terms of 
interacting spinless holes and spin waves in order to make it similar to that of the polaron problem. Then in the self-consistent Born approximation 
the selfenergy may be calculated (\cite {Mart}, \cite{SR})
$$
\Sigma ({\bf k}, \omega) = \sum_q f({\bf k},{\bf q}) G({\bf k-q},\omega - E_q)
$$
and the result is an integral equation for the Green's function of the form
\begin{equation}\label{integr}
G(k,\omega) =\frac 1{\omega-\sum_q f(k,q)G(k-q,\omega-E_q) }
\end{equation}
  where $f(k,q)$ contains information about the coupling of the holes to the spin
excitations and is of the order of $t^2$ , $t$ being the hopping rate  of the hole;  $E_q$ is the energy of the spin excitations, 
which is of order of $J$, the antiferromagnetic exchange energy.
 This self-consistent integral equation contains all the information about the hole spectrum ; numerical analysis  has shown, that it consists  
of a broad band of incoherent states and near to its bottom - a peak in the density of states (DOS), corresponding to a quasiparticle state 
(see e.g.\cite{Mart}).

It is often claimed that the ${\bf k}$-dependence is not too important \cite{Mohan}, \cite{SR}. Then it is worth to analyze 
thoroughly the ${\bf k}$-independent  case ( the Ising limit), when (\ref{integr}) take a simpler form:
\begin{equation}\label{GG}
 G(\omega) = \frac 1{ \omega - t^2 G(\omega - J)}.
 \end{equation}
(We have absorbed a factor $z$ (the coordination number) into $t^2$)

 This functional equation (\ref{GG}) is investigated in the literature in relation with the polaron problem \cite{Pines}, \cite{Appel}, 
the problem of the hole motion, described above, but also in other physical problems. We mention  another example here - the one-dimensional 
anisotropic ferromagnetic chain, which is studied in relation with
 experiments on the magnetic salt $CoCl_2 . 2H_2O $ \cite{Fog}, \cite{Torr}.

The resulting DOS consists of a series of sharp delta functions, as shown i.e in \cite{Kane} by numerical calculations. This picture is in 
contrast with the smooth DOS of an incoherent band for $J=0$ found in \cite{BR} and it is not clear how the limit $J\to 0$ can be obtained.

 In order to understand the behavior of the system it is useful to realize, that the motion of the hole creates "strings" of overturned spins; 
their field confine the hole, thus leading to selftrapped states centered at the original 
hole position\cite{68}, \cite{Hirsch}, \cite{Trug}, \cite{Sigg}.

 Here we present an approximated treatment of eq. (\ref{GG}) suitable for small values of $J$ and which allows us to consider the limit $J\to0$. 
We transform the functional equation into a differential equation and we show that the solutions of the latter present a picture of the DOS in 
agreement with the existing numerical calculations.

The paper is organized as follows. First we discuss the existing solutions of eq. (\ref{GG}), taking as a model the investigations on magnetic salts.
Then we transform the functional equation into a differential equation and we solve it. At the end we analyze the resulting DOS and outline the changes 
induced by the interaction $J$. We show that our solutions are in line with the previous calculations.

\section{Discussion of the existing analytical solutions}

In order to outline the simplicity of the method, proposed here, we first recall the exact solution, following in the main the calculation for the 
case of magnetic salts. \cite{Fog}, \cite{rev}.

First of all notice that the functional equation (\ref{GG}) is equivalent to the continued fraction :
$$
G = \frac 1 {\displaystyle \omega - \frac{t^2}{\displaystyle \omega - J - \frac{t^2}{\omega - 2J -\,.\,.\,.}\,\,} }\, ,
$$
which in turn can be thought of as solution of the series of difference equations:
\begin{equation}\label{coupled}
c_n(\omega - nJ) + t(c_{n+1} +c_{n-1})=0
\end{equation}
for n = 0, 1 , 2 , . . .
Indeed, from (\ref{coupled}) one obtains
$$
c_0= \frac {t c_{-1} }{\displaystyle \omega - \frac{t^2}{\displaystyle \omega - J - \frac{t^2}{\omega - 2J -\,.\,.\,.}\,} }\, .
$$
The spectrum may be found from the boundary condition $c_{-1} = 0 $.

 Compare now to the case of magnetic salts \cite{rev}. The Heisenberg Hamiltonian with a magnetic field  $H_0$ is simplified to reflect the 
main properties of the system. So, one starts from the eigenstates of the Ising:
$$
H^0 = -2 J^z\sum_{i=1}^N S^z_i S^z_{i+1}+ \gamma H_0 \sum_{i=1}^N  S_i^z \, ,
$$
which consist of single clusters of $n$-adjacent spin deviations with respect to the ferromagnetic ground state, with selfenergies
 \begin{equation}\label{Zeeman}
E^{(0)}_n = 2 J^z + n\gamma H_0,
 \end{equation}
and constructs Bloch functions. In this basis is calculated the perturbation due to the anisotropic term, which is the most important in the case of 
ferromagnetic salts:
$$
H^a = -J^a\sum_{i=1}^N (S^+_i S^+_{i+1} + H.C.)\equiv H^{pert}.
$$
For the coefficients in the perturbation series one obtains the same type of coupled equations as (\ref{coupled})
 the only difference being due to the fact, that the perturbation $H^{pert}$ couples states with spin deviation differing by 2 \cite{Fog}:

$$
n\gamma H_0 - J^a \cos(ka)(c_{n+2}+c_{n-2}) = (E- 2J^z) c_{n}
$$
with $a$ - the interatomic distance and $ -\frac \pi a \leq k \leq \frac\pi a$.

The correspondence between the two cases, discussed above, can be expressed as follows:
$$
(E- 2J^z) \to \omega ; \quad\gamma H_0 \to J ; \quad J^a \cos(ka) \to t .
$$

For the simplest case  $J=0$ (or in the absence of magnetic field, when one considers the magnetic salts)
 (\ref{coupled}) may be solved taking ${\displaystyle \frac {c_{n+1}}{c_n}= \frac{c_n}{c_{n-1}} = T}$.
The "transfer matrix" $T$ obeys the equation:
$$
T\omega + t(T^2 +1)=0
$$
and we have
$$
T=\left( - \omega + \sqrt{\omega^2 - 4 t^2}\right)/2t
$$
We obtain continuous spectrum in the region  $ -2t\leq  \omega \leq 2t $ . This corresponds to the semispherical DOS found in \cite{BR}. 
Let us put $\omega = 2t \cos \lambda $,  where $\lambda\in (0, \frac\pi 2)$.
For $T$ we have two solutions: $T = e^{\pm i\lambda}$, from which we can construct the expressions
for $c_n$, corresponding to the appropriate boundary condition : $c_n = A\sin((n+1)\lambda) $.

Now, let us consider the solution of (\ref{coupled}) in the general case. When we realize that
the same recurrence relation satisfy the Bessel's functions, we can write this solution at once:
\begin{equation}
c_n = A J_{\left(\frac{\omega}J -n\right)/2} \left( \frac {|t|} J\right) .
\end{equation}
The boundary condition is then
$$
J_{\left(\frac{\omega}J +1\right)/2} \left( \frac {|t|} J\right) = 0 \,.
$$
 The spectrum can be found as the zeros of the Bessel functions are tabulated
(see i.e. \cite{zero}).

It is possible to obtain more transparent expressions using the appropriate double asymptotic
 expansion  of the Bessel function in both order and argument \cite{Fog} ,
\cite{math}:
$$
J_\mu (b\mu) \approx \sqrt{\frac 2{\pi\mu\tan\alpha}}{ \cos[\mu(\tan\alpha - \alpha)- \pi/4]\, ,
+O(\mu^{-1/5}) }
$$
where $ b = 1/\cos \alpha , 0<\alpha < \pi/2$.

Setting $\omega = 2t \cos \lambda $ as earlier, we have
$$
b = \frac{2 t}{\omega} =\frac 1{\cos\lambda} = \frac 1{\cos\alpha}
$$
i.e $\lambda =\alpha$
and the zeros of the Bessel functions can be calculated from the condition:
$$
\sin\lambda- \lambda \cos \lambda = \pi(\nu +\frac 3 4) J /2t\, ,
$$
where $\nu = 0, 1, 2...$. From this formula the spectrum can be calculated numerically.

Another type of approximation to the recurrence equation(\ref{coupled})
consists of considering the coefficients $c_n$ as slowly varying functions of $n$, which
is justified in the limit of large $n$
and is on the same lines as the calculation in \cite{68}.
Replacing $c_n$ by a continuous function $c(n)$ and expanding $c(n\pm1)$ about $n$, we obtain the following equation:
\begin{equation}
c(n)(\omega - nJ -2t) +  tc''(n) = 0
\end{equation}
which may be regarded as a one dimensional Schr\"odinger equation, describing the motion of a particle of mass $1/2t$ in the linear ("wedge") 
potential $U(n) = nJ$. The boundary condition $c(-1) = 0$ is equivalent to the introduction of an infinite potential wall. This problem is well 
known to yield solutions in terms of Airy-functions. In WKB approximation the Bohr-Sommerfeld condition reads:
$$
 \int p(x) dx =\int_0^{x_0}  \sqrt{(\omega -2t +J -xJ)/t}\, dx=
 \pi(\nu +3 /4) , \quad \nu = 0, 1, 2 ...,
 $$
 where $x = n+1$ , \, $x_0 = (\omega -2t +J )/J $
 and one obtains the spectrum
 $$
 \omega -2t +J =   (3\pi(\nu +3/ 4)/2)^{2/3} t^{1/3} J^{2/3} .
 $$

The spectrum, found here, consisting of discrete levels, should merge to form a continuous band, when $J\to 0$. But this fact cannot be expressed 
analytically,  the limit $J\to 0$ being singular.

\section{A new approach to the functional equation}

Let us write the equation for the self-energy, corresponding to eq. (\ref{GG}):
\begin{equation}
\Sigma (\omega) = \frac{t^2}{\omega- J - \Sigma(\omega - J )} \,.
\end{equation}

For small $J$ we can expand the function $\Sigma$\,:
\begin{equation}
\Sigma (\omega) = \frac{t^2}{\omega- J  - [\Sigma(\omega)- J \Sigma'(\omega)
+\frac1 2( J )^2\Sigma''(\omega)+... ]} \, ,
\end{equation}
or
\begin{equation}\label{sec}
\Sigma(\omega- J  -\Sigma) + J \Sigma'\Sigma -\frac1 2( J )^2\Sigma''\Sigma + ...
\approx t^2
\end{equation}
When $J=0$ the resulting equation
$$
\Sigma(\omega - \Sigma) =t^2
$$
has the well known solution
\begin{equation}\label{Sigma}
\Sigma =(\omega \pm \sqrt{\omega^2 - 4t^2})/2
\end{equation}
which gives a semispherical density of states (DOS) in the region $ |\omega| \leq 2t$.

 The differential equation (\ref{sec}) cannot be integrated analytically. Nor is it possible
 to treat the terms $\sim J$ as a perturbation; so, in order to find an appropriate solution
 we proceed as follows:

  We will transform first (\ref{sec}) into
  a higher order equation,
 \begin{equation}
\Sigma'(\omega- J  -\Sigma)+\Sigma(1-\Sigma') + J (\Sigma')^2 +
 J \Sigma''\Sigma -\frac1 2( J )^2\Sigma''\Sigma'-....=0 \, ,
 \end{equation}
having in mind, that only part of the solutions of the
 latter are also solutions of (\ref{sec}). Then we have to choose the integration constants
 as to reproduce the correct result for the equation of lower order.

 We are interested in the solution for the energy region of the quasi-band states,
 as we want to know how the DOS, generated from (\ref{Sigma}), changes,
 when one includes
 a small interaction $J$. We look for a solution, which is sufficiently smooth
 \footnote{Because of this approximation details of the DOS structure near the band edges are lost}.
  So we neglect the terms $\sim J\Sigma''$ and obtain the following equation
 ($x\equiv \omega-J $ and $ y \equiv \Sigma$):
 \begin{equation}\label{eq}
 Jy'^2 +y = (2y -x)y' \, .
 \end{equation}
 We believe this equation contains the essential part of the information needed to analyse
 the change of the DOS in the band. In the {\bf Appendix} we outline the main steps of
 the calculation. The solution is obtained in parametric form:
 \begin{equation}\label{xy}
 y =\sqrt{D\frac p{p-1}}+ \frac J 2\, p-
 \frac J 2 \,\,\sqrt{\frac p{1-p}}\arctan \sqrt{\frac p{1-p}}
\end{equation}
$$
 x =\frac{2p-1}p \sqrt{D\frac p{p-1}}- \frac J 2 -
  \frac J 2\,\, \frac{2p-1}p \sqrt{\frac p{1-p}}\arctan \sqrt{\frac p{1-p}},
 $$
 where $D$ is an integration constant. In order to reproduce the known results without $J$,
 we have to choose $D = t^2$.
Indeed, let us take $J=0$, we have then:
$$
y^2 = D\frac p{p-1} \quad {\mbox and} \quad x = \frac{2p-1} p \,y
$$
whereby, getting off the parameter $p$ , we obtain  the direct relation between $x$ and $y$:
$$
x = \frac D y + y
$$
and we have again (\ref{Sigma}).

Another useful relation reads
\begin{equation}\label{use}
p = \frac 1{2J} [\,2y -x\mp\sqrt{(2y-x)^2 - 4Jy}\,].
\end{equation}
(The sign is to be chosen properly, as to reproduce at $J\to 0$ the starting equation
$ y'=y/(2y-x)$ ).

 The next step is to transform these relations in some formulae, which enables us to
 analyze the change in the DOS under the influence of $J$.

We rewrite the second equation in (\ref{xy}) in the form:
$$
 \tilde{x} \equiv x+\frac J 2 =\frac{2p-1}p \sqrt\frac p{p-1} \,\tilde{t}
$$
where
$$
\tilde{t}= \sqrt D - i\frac J 2  \arctan  \sqrt\frac p{1-p}
$$
(remember, that $\sqrt D \equiv t $).
Then we introduce a parameter $q$:
$$
\frac{(2p-1)^2}{p(p-1)} = \frac{\tilde{x}^2}{\tilde{t}^2}\equiv -q
$$
and find, making use of (\ref{use}):

$$
y =\frac 1 2 \left[x+J +\sqrt{\frac{[x(q+4)+2J]^2}{q(4+q)}+2Jx + J^2}  \right]
=
 \frac 1 2 \left[ \tilde{x} \left(1 + \sqrt\frac{4+q}q \right )
+\frac J 2 \left(1+\sqrt\frac q{4+q}\right )\right]
$$
 Then the parameter $\tilde{t}$ is obtained from the equation
 \begin{equation}\label{tt}
\tilde{t} = t +
\frac J 2 Arth \frac{ \tilde{x}+ \sqrt{ \tilde{x}^2- 4\tilde{t}^2}}{2\tilde{t}}
 \end{equation}
and the solution  for $y$ is:
\begin {equation}\label{y-x}
y = \frac{ \tilde{x}+ \sqrt{ \tilde{x}^2- 4\tilde{t}^2}}{2}
\left( 1+ \frac J 2 \frac 1{\sqrt{ \tilde{x}^2- 4\tilde{t}^2}} \right)
\end{equation}

It is clear, that the band width is now determined by the value of $ \tilde{t}$, which is
function of $x$ to be calculated from (\ref{tt}). This transcendental equation  can
be solved numerically, but several properties may be revealed by analytical study of the results.

\section{Analysis of the solution}
We will still use for convenience the variable $ \tilde x =\omega - J/2  $ (a shifted energy).
From (\ref{y-x}) we have  the self-energy
\begin {equation}
\Sigma = \frac{ \tilde{x}+ \sqrt{ \tilde{x}^2- 4\tilde{t}^2}}{2}
\left( 1+ \frac J 2 \frac 1{\sqrt{ \tilde{x}^2- 4\tilde{t}^2}} \right)
\end{equation}
and we can calculate the Green's function
$$
 G = \frac 1 {\tilde x +\frac J 2  - \Sigma}
$$
in terms of the parameter $\tilde t$:
\begin{equation}
G= \frac {(\tilde x ^2 - 4\tilde t^2) (\tilde x + \frac J 2) 
+ \sqrt{\tilde x ^2 - 4\tilde t^2} (\tilde x ^2 - 4\tilde t^2 + \frac{J\tilde x} 2)} {2\tilde t^2 (\tilde x ^2 - 4\tilde t^2 -\frac {J^2}{8})}
\end{equation}
The parameter $\tilde t$ plays in this formula a role similar to that of the hopping rate $t$ in the absence of the interaction $J$. 
In the latter case $t$  determines the width  of the incoherent band ($4t$). Now $\tilde t$ depends on  $\tilde x $ (on the energy) and 
it is not obvious how the band width changes.

\begin{figure}
\centering
\rotatebox{-90}{
  \includegraphics[width=10cm]{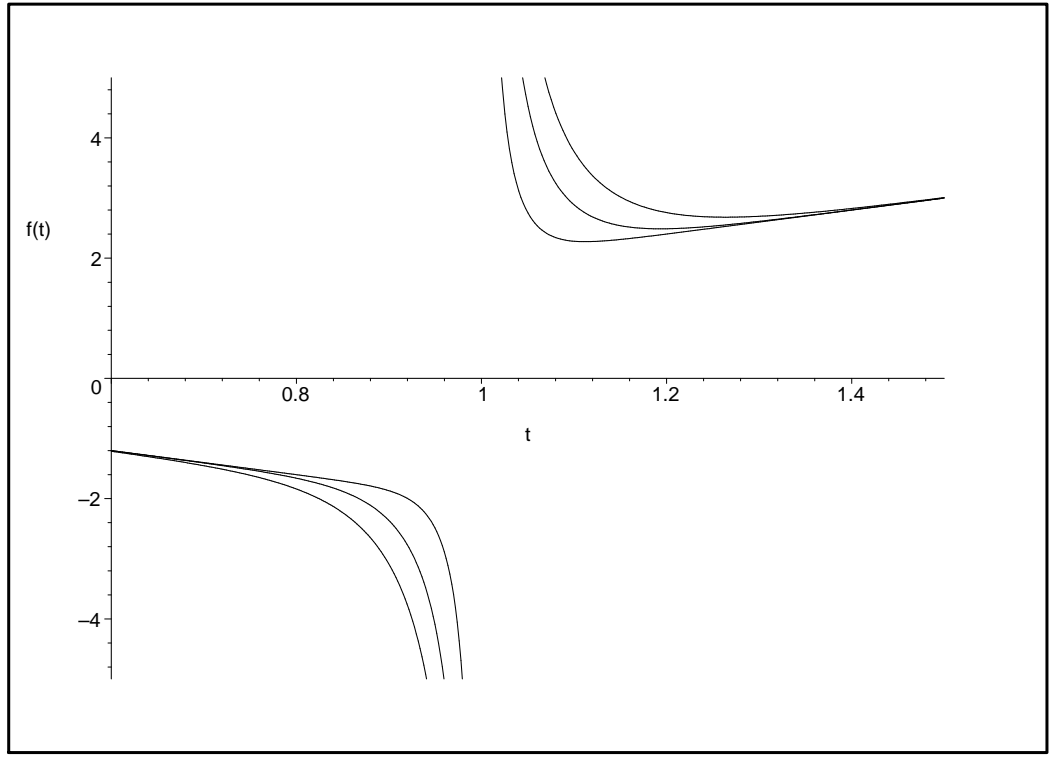}}\\
  \caption{Plot of the function $f(\tilde t)$ (the symbol t on the plot stays for $\tilde t $ and the hopping rate $t=1$) for  
3 different values of $J$ :0,2 ;0,4 and 0,6.}\label{fig}
\end{figure}

The relation to determine $\tilde t$ (\ref{tt}) may be rewritten in the form
\begin{equation}\label{t-x}
\tilde x = 2\tilde t \, \coth\left(\frac 4 J (\tilde t - t)\right) \equiv f(\tilde t),
\end{equation}
so, the function $\tilde t(\tilde x)$ can  be found graphically using the plot of $  f(\tilde t)$ (Figure \ref{fig}).

%%%%%%%%%%%%%%%%%%%%%%%%%%%%%%%%%%%%%%%%%%%%

%%%%%%%%%%%%%%%%%%%%%%%%%%%%%%%%%%%%%%%%%%%%%%%%%

First we consider the region above the top of the band. For high energies (large $\tilde x >0$) solution is obtained from the curve  $f(\tilde t)$ 
above the jump at $\tilde t =t (t=1)$ and we have  $\tilde t> 1$.
The physically meaningful  solution for $\tilde t(\tilde x)$ is obtained on the left branch of the curve , where $\tilde t(\tilde x) \to t $ 
as $ \tilde x \to \infty$.

Once $\tilde t$ is known, one can calculate the Green's function. In this region of $\tilde x$  our solution satisfies $\tilde x > 2 \tilde t$ 
and $G$  is real. We may use the following expression for it:
\begin{equation}\label{Greal}
 G = \frac {\tanh \left(\frac 2 J(\tilde t - t) \right)}{ \tilde t +
  \frac J 4 \sinh \left(\frac 4 J(\tilde t - t) \right)}
\end{equation}

The function$f(\tilde t)$ - (\ref{t-x}) - has a minimum in this region and the corresponding value of $\tilde t , \,\tilde t_m$ \, is to be 
evaluated from the relation :
$$
\sinh\left((8/J)(\tilde t_m -t)\right) = (8/J) \tilde t_m .
$$

Below the minimum of $f$ real solutions of the transcendental equation (\ref{t-x}) do not exist, so it represents the upper limit (the top) of the band.
We have calculated the minimum value of $f$ (or the top of the band $\tilde x _{top} $) for  3 different values of the exchange 
interaction $J$( see Table 1).\rule{0pt}{5mm}

\begin{center}
\begin{tabular}{|l|l|l|l|}
\multicolumn{4}{c}{Table 1}\\
\hline
$J/t$  & $0,2$ & $0,4$ &$0,8$\\
\hline
$\tilde x_{top}/t$ &2,275 &2,489 & 2,8637\\
\hline
$\tilde t_{top}/t$& 1,11222 & 1,1934 & 1,3285\\
\hline
$\tilde x_q/t$ & -1,649 & -1,40 &-1,23\\
\hline
\end{tabular}
\end{center}

\rule{0pt}{5mm}
We notice the  almost linear increase of the top of the band with $J$ beginning from its value of $2t$ at $J=0$.

For $\tilde x < f_{min}$ we have to calculate the complex $\tilde t= t_1 + it_2 $ and then find the complex Green's function which gives the  
spectral function (and the DOS in the band).
From (\ref{t-x} ) we obtain the following system for the real and imaginary part of $\tilde t$:
\begin{equation}\label{compl 1}
%\frac {\tilde x}2 \frac{16} J =
\frac4 J \tilde x =\frac {2 \eta}{\sin(2\eta )} B ;
\end{equation}
\begin{equation}\label{compl 2}
\frac {2 \eta}{\sin(2\eta )} =\frac {2 \xi +\frac{8} J}{\sin(2\xi )}
\end{equation}
where the following notations are used:
$$
\xi = \frac 4 J t_1 - \frac 4 J t \quad, \quad \eta = \frac 4 J t_2 \, \quad ,\quad
B = \cosh(2\xi) +\cos(2\eta).
$$
In terms of these variables we may write the Green's function in the form:
\begin{equation}
G\, = \, \frac{8}J \frac{\cosh(\xi - i \eta)}{[\cosh(\xi - i \eta) - B]\left[{\displaystyle 1- \frac{2\eta}{sin(2\eta)}\cosh(\xi - i \eta)}\right]} \, .
\end{equation}

 We  study analytically $G$ in the center of the unperturbed band ($\tilde x =0$) and in its vicinity.
 For $\tilde x = 0, \xi = 0 $ and $ 2\eta =\pi $ .
 Let us define
 $$
2\xi = \delta ; \quad  2\eta = \pi + \Delta .
 $$
and consider the case $ \delta\ll 1$ and $ \Delta \ll 1$. From (\ref{compl 1}) and (\ref{compl 2}) we find
\begin{equation}
\delta \approx \frac {\tilde x}{1 + (J\pi/8)^2} ;\quad  \Delta \approx - \frac{\delta \pi}{2\delta + 8/J}
\end{equation}
and for the Green's function
\begin{equation}
G= \frac {8}J \, \frac 1 {\displaystyle 2-\pi -i\pi\frac{\delta}{\Delta}} .
\end{equation}
As $\delta/\Delta \approx - 8/J\pi$ roughly, we obtain for the real and imaginary part of $G$ (the latter gives the DOS) in 
the vicinity of $\tilde x = 0$
\begin{equation}
Re\, G = \frac {(J/8)(2-\pi)}{1+\left[ (2-\pi)J/8\right]^2} \, ,
\quad  Im \, G = \frac {-1}{1+\left[ (2-\pi)J/8\right]^2}.
\end{equation}

Another important feature of the calculated Green's function is it's behavior in the region of the lower part of the band. Consider now 
the negative values of the energy (of $\tilde x$ ) . At $\tilde x \to -\infty $
we find again $\tilde t \to t$ , but $\tilde t < t$ (on the left side of the jump of the function $f(\tilde t)$ ), as it is expected.
The ratio $|\tilde x/2\tilde t|>1 $ , so we have real values of the Green's function and we are outside the band.

On the other hand, as $\sinh\left(\frac 4 J(\tilde t - t) \right) <0$ for $\tilde x<0$,  we see from (\ref{Greal}) 
that the Green"s function has a pole. Hence, we obtain  a quasi-particle state. The corresponding energy is calculated from the condition:
\begin{equation}
(4/J) \tilde t = -\sinh\left((4/J)(\tilde t - t)\right)
\end{equation}
and making use of (\ref{t-x}):
$$
\frac{\tilde x}{2 \tilde t} \,=\, \frac {\sqrt {1 + \sinh^2\left(( 4/ J)(\tilde t - t) \right)}}
{\sinh\left( (4 /J)(\tilde t - t) \right)} .
$$
We find
\begin{equation}
\sqrt {\tilde x ^2 - 4 \tilde t^2(\tilde x)} = J/2  %\frac J 2
\end{equation}
for the energy of the quasiparticle level.
The calculated  energies $\tilde x_q$ of this level
for 3 different values of the parameter $J$  are given in Table 1.

For negative energy smaller in absolute value we arrive at the point, where the condition
$2\tilde t \approx \tilde x$ is satisfied.  (For small $J$ this point is just next to the quasiparticle level). The latter condition is the 
bound of the real values of $\coth\left(\frac 4 J (\tilde t - t)\right)$. So , we are at the lower limit (the bottom) of the band.

\section{Conclusions}.
We developed here a different method for solving the functional equation (\ref{GG}) by transforming it into a differential equation. 
This approach may be useful when the exchange interaction $J$ is not large. Under this approximation the region of sharp peaks in the DOS merge 
into a smooth band. We can see how the top and the bottom of the band  change with $J$ - they are shifted to higher energies. numerical results 
obtained for the more general equation (\ref{integr}).
Similar behavior is obtained as a result of numerical analysis of the more general equation (\ref{integr})(see e.g. \cite{Mohan},\cite{SR}).

  Also a bound level (quasiparticle state) immediately under the incoherent band is found, in agreement with previous numerical calculations \cite{SR}.

 So we have a picture of the DOS, which is very close to that known from the literature.

We also obtained an explicit expression for the Green's function near the center of the band.

We believe, that the proposed method of treating  the functional equation can be used as a complementary tool to study the DOS, 
generated by the self-consistency equation (\ref{GG}).

Numerical calculations of the DOS in the entire band are in course.

\section{Appendix}

Here we write down the main steps in finding the solution of (\ref{eq}) of the text.

First we label $y'\equiv p$ and we write it in the form:
\begin{equation}\label{x-p}
x = 2y-Jp -\frac y p.
\end{equation}
We then differentiate to obtain:
$$
dx = 2 dy - J dp -\frac 1 p dy + \frac y{p^2}dp = \frac{dy} p
$$
Then we are at the equation
$$
\frac{dy}{dp} - \frac y {2p(1-p)} + J\frac p{2(1-p)} = 0
$$
and its solution
$$
y(p) = \sqrt\frac p{1-p}\left[C -\frac J 2 \left( \arctan\sqrt\frac p{1-p} -
\sqrt {p(1-p)}\right)\right]
$$
Returning to (\ref{x-p}) we have:
$$
x(p) = \frac{2p-1}p\, \sqrt\frac p{1-p}\left[C -\frac J 2 \arctan\sqrt\frac p{1-p}\right]
-\frac J 2.
$$
Redefining the integration constant, we can write
$C\sqrt\frac p{1-p} \equiv \sqrt{D\frac p{p-1}}$ and then system
\begin{equation}
y(p)=\sqrt{D\frac p{p-1}}+ \frac J 2\, p-
 \frac J 2 \,\,\sqrt{\frac p{1-p}}\arctan \sqrt{\frac p{1-p}}
\end{equation}
$$
x =\frac{2p-1}p \sqrt{D\frac p{p-1}}- \frac J 2 -
  \frac J 2\,\, \frac{2p-1}p \sqrt{\frac p{1-p}}\arctan \sqrt{\frac p{1-p}}
$$
define our parametric solution of (\ref{eq}).

Without $J$ the solution is readily obtained, writing (\ref{eq}) as $2yy' = xy' + y$ and
integrating:
$$
y^2 + C_1 = xy ;
$$
here the integration constant  $C_1$ is identified as $ C_1=D$.


\begin{thebibliography}{tbds}
\bibitem {BR}
W.F. Brinkman and T.M. Rice Phys.Rev. B {\bf 2}, 1324 (1970).
\bibitem{Mart}
G. Martinez and P.Horsch, Phys. Rev.B {\bf 44}, 317-31 (1991)
\bibitem {SR}
S.Schmitt-Rink, C.M. Varma and A.E. Ruckenstein, Phys.Rev.Letters {\bf 60}, 2793-96 (1988).

\bibitem{Szcz}
K.J. von Szczepancki, P.Horsch, W.Stephan and M.Ziegler, Phys.Rev. B {\bf 41}, 2017-29 (1990).

\bibitem{Kane}
C.L. Kane, P.A.Lee, N.Read, Phys.Rev.B, {\bf39},6880-97 (1989).
\bibitem{Mohan}
M.M. Mohan, J.Phys.:Condens. Matter {\bf 3}, 4307-12 (1991).

\bibitem{Pines}
D.Pines, Polarons and Excitons, 1962.
\bibitem{Appel}
J.Appel in Solid State Physics. Vol.21 (Seitz and Turnbull 1968).

\bibitem{Fog}
H.C. Fogedby and H.H. Jensen, Phys. Rev. B {\bf6}, 3444-6 (1972).
\bibitem{Torr}
J.B. Torrance Jr, M.Tinkham, Phys. Rev {\bf 187}, 587, 595 (1969).
\bibitem{rev}
H.C.Fogedby, Lecture notes in physics {\bf 131},(Springer-Verlag, Berlin, Heidelberg,
New York 1980).

\bibitem{68}
L.N.Bulaevski, E.L.Nagaev and D.I.Khomski, Sov.Phys.-JETP {\bf 27}, 837 (1968).
\bibitem{Hirsch}
J.E.Hirsch, Phys. Rev. Lett. {\bf 59}, 228 (1987).
\bibitem{Trug}
S.A. Trugman, Phys. Rev B {\bf 37}, 1597 (1988).
\bibitem{Sigg}
B.I.Shraiman and E.D.Siggia, Phys.Rev.Letters {\bf60}, 740 (1988).

\bibitem{zero}
E. Jahnke, F.Emde, F.L$\ddot o$sch, Tafeln H$\ddot o$herer Funktionen, (B.G.Teubner Verlagsgeselschaft, Stuttgart, 1960)
\bibitem{math}
R.Courant and D. Hilbert, Methods of Mathematical Physics (Interscience, New York, 1966).

\end{thebibliography}
\end{document}